\documentclass[pra,twocolumn,preprintnumbers,superscriptaddress]{revtex4} 

\usepackage{graphicx}
\usepackage{bm}
\usepackage{amsmath}
\usepackage{amssymb}
\usepackage{amsfonts}
\usepackage{fancyhdr}
\usepackage{slashed}
\usepackage{latexsym,epsfig,bbm}
\usepackage[colorlinks=true,urlcolor=blue,citecolor=blue]{hyperref}

\usepackage{color}
\definecolor{blue}{rgb}{0,0.2,1}

\definecolor{red}{rgb}{0.9,0,0}

\usepackage{amsthm}
 
\theoremstyle{definition}
\newtheorem{definition}{Definition}

\begin{document}
	
\title{Vulnerability of quantum classification to adversarial perturbations}
\author{Nana Liu}
\email{nana.liu@quantumlah.org}
\affiliation{John Hopcroft Center for Computer Science, Shanghai Jiao Tong University, Shanghai 200240}
\author{Peter Wittek}
\affiliation{University of Toronto, M5S 3E6 Toronto, Canada}
\affiliation{Creative Destruction Lab, M5S 3E6 Toronto, Canada}
\affiliation{Vector Institute for Artificial Intelligence, M5G 1M1 Toronto, Canada}
\affiliation{Perimeter Institute for Theoretical Physics, N2L 2Y5 Waterloo, Canada}
\begin{abstract}
High-dimensional quantum systems are vital for quantum technologies and are essential in demonstrating practical quantum advantage in quantum computing, simulation and sensing. Since  dimensionality grows exponentially with the number of qubits, the potential power of noisy intermediate-scale quantum (NISQ) devices over classical resources also stems from entangled states in high dimensions. An important family of quantum protocols that can take advantage of high-dimensional Hilbert space are classification tasks. These include quantum machine learning algorithms, witnesses in quantum information processing and certain decision problems. 
However, due to counter-intuitive geometrical properties emergent in high dimensions, classification problems are vulnerable to adversarial attacks. We demonstrate that the amount of perturbation needed for an adversary to induce a misclassification scales inversely with dimensionality. This is shown to be a fundamental feature independent of the details of the classification protocol. Furthermore, this leads to a trade-off between the security of the classification algorithm against adversarial attacks and quantum advantages we expect for high-dimensional problems. In fact, protection against these adversarial attacks require extra resources that scale at least polynomially with the Hilbert space dimension of the system, which can erase any significant quantum advantage that we might expect from a quantum protocol.
This has wide-ranging implications in the use of both near-term and future quantum technologies for classification.
%We also show corresponding results for continuous variable systems when the system has infinite dimensional degrees of freedom. To mention only as paragraph in the discussion/outlook
\end{abstract}

\maketitle

Quantum technologies promise exciting advantages in quantum computation \cite{nielsen2000quantum}, simulation \cite{cirac2012goals}, metrology \cite{giovannetti2011advances} and cryptography \cite{ekert1991quantum}. Even while large-scale and fault tolerant quantum technologies currently remain out of reach, noisy intermediate-scale quantum (NISQ) devices \cite{preskill2018quantum} hope to deliver advantages over classical systems in the near-term. Most of these protocols exploit not only unique quantum characteristics like entanglement and superposition, but also high-dimensional Hilbert spaces. Without the latter, no sizeable quantum advantages in either computation, simulation or sensing are expected. 

One important class of tasks where high-dimensional Hilbert spaces may be advantageous are classification problems. To correctly categorise an object belongs to one of the most common and basic questions asked in science. In the quantum setting, classification problems can appear mainly in one of two ways. Firstly, a quantum protocol can be used for classification problems with classical data in order to gain a quantum advantage in speed or precision. Many quantum-enhanced machine learning algorithms are of this type \cite{biamonte2017quantum}. Alternatively, one can classify quantum states or processes themselves, in terms of entanglement \cite{ma2018transforming}, phases \cite{carrasquilla2017machine} or other many-body behaviour. Quantum learning protocols also belong to the latter category \cite{monras2017inductive}. 

From these examples, it therefore appears that advantages for classification tasks will become more pronounced as Hilbert-space dimension grows, at least in the absence of noise. However, as we will see, this is no longer true when the classification protocols are subject to adversarial perturbations. These are small, often hard-to-detect perturbations of the object to be classified which give rise to deliberate misclassification. This is highly relevant for many classification problems.  In the machine learning context in particular, security breaches in the algorithm are not only desirable for adversarial parties, but also made possible as data used for classification are often shared amongst multiple, possibly untrusted parties \cite{huang2011adversarial}. Recent findings in machine learning suggests that even highly successful classification algorithms can be very vulnerable to adversarial perturbations if the dimension of the data is high enough, such as high-resolution image data \cite{szegedy2013intriguing}. Some quantum machine learning algorithms that resist certain kinds of adversarial attacks have also been recently developed \cite{wiebe2018hardening}. However, it is yet unknown what the fundamental limits to adversarial robustness are for quantum classification problems in general. 

In this paper, we demonstrate that a perturbation by an amount scaling inversely with the dimension of the quantum system to be classified is sufficient to induce a misclassification. Amazingly, this is a fundamental feature of quantum classification originating from a purely geometrical property of high-dimensional spaces, known as the concentration of measure phenomenon \cite{ledoux2001concentration}. It is independent of the specifics of any particular classification protocol. Furthermore, detection of these small perturbations by existing efficient certification protocols for quantum systems cannot be efficient. For classification problems, otherwise efficient certification protocols must now require resources scaling polynomially with dimensionality. This is an exponential resource cost in the corresponding number of qubits and can thus potentially erase any key quantum advantage from using high-dimensional quantum systems in the first place. 

This result presents a fundamental trade-off between the resources for preserving the security of the quantum system and resource advantages for classification protocols that exploit high-dimensional Hilbert space. Classification protocol that classify either raw quantum data or classical data embedded into quantum states will be subject to this limitation. Thus, this leads to widespread implications for many quantum classification tasks including, quantum machine learning algorithms, some witnesses in quantum information amongst others. To gain a more intuitive understanding of this result, we begin with examining what is special about classification problems, whose output belongs to a countable set. We compare this to quantum computational algorithms or other quantum protocols whose output can take on any continuous value. 

Suppose we have two parties, Alice and Bob. Alice wants to perform a classification task on an input quantum state $\sigma$, which is selected from the set $\Sigma$. The classification device Alice uses implements a function $v: \Sigma \rightarrow \Re$. This function can be written as  $v(\sigma)=\text{Tr}(\mathcal{O}\Lambda(\sigma))$, where $\Lambda$ is a CPTP map performed by the device and $\mathcal{O}$ is an observable measured at the output of the device. For a classification problem, the final answer needs to be constrained to a discrete set of values and these are known as class labels. Thus $v(\sigma)$ can be converted into discrete values $h(\sigma)$ by using some thresholding function, e.g. $h(\sigma)=\text{sign}(v(\sigma))$ for binary classification. 

However, Bob may be a possibly adversarial party who can induce a small perturbation about $\sigma$ that goes undetected by Alice. This means $\sigma$ is modified into a state $\rho$ where Alice is able to detect when $F(\sigma, \rho)<1-\chi$ but unable to detect when $F(\sigma, \rho)\geq 1-\chi$. Then it can be shown, by applying Uhlmann's theorem, that $|v(\rho)-v(\sigma)|$ is bounded above by a quantity proportional to $\sqrt{\chi}$ (see Appendix~\ref{app:uhlmann}). Thus a small $\chi$, which corresponds to a large overlap between $\sigma$ and $\rho$, can result in a small difference $\mathcal{O}(\chi)$ in the final real-number output of the quantum device. %\footnote{This is reminiscent of include Preskill's result in words ($P_{fail}\leq \mathcal{O}(\epsilon^2)$.}

Although this might suggest that quantum protocols may in some cases be robust against perturbations of the input state, adversarial or otherwise, this inference does not necessarily hold for classification problems. Classification problems care about the discrete class label $h(\sigma)$ rather than $v(\sigma)$ itself. Thus if $v(\sigma)$ lies very close to the decision boundary for $h$, even a small $\chi$ with corresponding small $|v(\rho)-v(\sigma)|$ can mean $\rho$ is classified differently to $\sigma$, i.e. $h(\sigma) \neq h(\rho)$. 

This observation hints that, for classification problems, it is really the concentration of points near a decision boundary that matters, which can be mathematically captured by the concentration of measure phenomenon. Although already widely-used in quantum information in the context of quantifying average quantum behaviour of random quantum states \cite{muller2011concentration} and related applications in cryptography \cite{hayden2004randomizing}, it has not yet been employed for classification problems in the quantum setting. Notably, in many situations, this concentration of measure phenomenon predicts a growing concentration of points near such a decision boundary as the dimensionality of the problem increases. We will see that this implies that higher-dimensional quantum classification problems are more vulnerable to adversarial perturbations. To formalize our result, let us first introduce the following definitions. 
\begin{definition}
Suppose we want to assign classification labels $s \in S$ to elements $\sigma \in \Sigma$ where $S$ is a countable set and $\sigma$ can be either classical or quantum data. $\sigma$ can refer to either states or processes. Functions $c:\Sigma \rightarrow S$, belong to the set $\mathcal{C}$ known as \textit{ground truth} if $c(\sigma)$ gives the true classification label for any $\sigma$. These functions are not generally directly accessible. Functions belonging to $\mathcal{H}$ known as the \textit{hypothesis class} are the accessible guesses to the ground truth.
%Thus if $g \in \mathcal{H}$, then $g(\sigma)$ is an approximation to $c(\sigma)$ for any $\sigma$.
%\footnote{We note that this general classification setting also applies to learning problems, which includes both supervised and unsupervised learning.}
\end{definition}
\begin{definition}
We select $\sigma$ from $\Sigma$ with respect to a probability measure $\mu$ and $\mu(\Sigma)=1$. Let $\mathcal{M}$ be the set of states that are misclassified by $h \in \mathcal{H}$, i.e, $\mathcal{M}=\{\sigma \in \Sigma| h(\sigma)\neq c(\sigma)\}$. Then we refer to $\mu(\mathcal{M})$ as the \textit{risk}, which is the probability that the selected $\sigma$ is misclassified by $h$.
\end{definition} 
\begin{definition}
Let there be metric defined over $\Sigma$ and the corresponding distance measure be denoted $D$. Then the $\epsilon-$expansion of subset $\tilde{\Sigma} \subseteq \Sigma$ is defined as the set 
\begin{align} \label{eq:epsilonexp}
    \tilde{\Sigma}_{\epsilon}=\{\sigma|D_{\text{min}}(\sigma, \tilde{\Sigma}) \leq \epsilon
    \},
\end{align}
where $D_{\text{min}}(\sigma_a, \tilde{\Sigma})$ is the infimum of the distance between $\sigma$ and any element of $\tilde{\Sigma}$. Then we call $\mu(\mathcal{M}_{\epsilon})$ the \textit{adversarial risk} under the perturbation of states $\sigma$ by distance $\epsilon$. This is equivalent to the probability that there is at least one state $\sigma'$ where $D(\sigma, \sigma')\leq \epsilon$ which is misclassified by $h$, i.e., $h(\sigma')\neq c(\sigma')$.
\end{definition}
\begin{definition}
For $\tilde{\Sigma}\subset \Sigma$ with distance measure $D$ and probability measure $\mu$, we can define the \textit{concentration function} $\alpha(\epsilon)$ as $\alpha(\epsilon) \equiv 1-\inf\{\mu(\tilde{\Sigma}_{\epsilon})|\mu(\tilde{\Sigma})\geq 1/2\}$. If $\tilde{\Sigma}$ is also equipped with a vector space with dimension $d$ and 
\begin{align} \label{eq:normallevy}
    \alpha(\epsilon)\leq l_1 e^{-l_2\epsilon^2d},
\end{align}
the corresponding space can be said to belong to the $(l_1,l_2)$-\textit{normal Levy family}, where $l_1, l_2>0$.
\end{definition}
Here $\alpha(\epsilon)$ quantifies the extent to which points in $\Sigma$ concentrate about the boundary lines of $\tilde{\Sigma} \subset \Sigma$. A small $\alpha(\epsilon)$ means a large concentration of points near such boundaries. For those spaces which belong to the normal Levy families, we see that this concentration near the boundary increase exponentially as the dimension $d$ of the space increases. This is the concentration of measure phenomenon.

We now focus our attention on $\Sigma=SU(d)$  where $SU(d)$ is the special unitary group. This choice will become clearer when we discuss our two scenarios. We can select $\sigma \in SU(d)$ randomly according to the Haar probability measure and our goal is to classify $\sigma$ where the risk is $\mu(\mathcal{M})$. Suppose $h(\sigma)=c(\sigma)$. However, if $\sigma$ is perturbed, for instance by an adversarial Bob, then the following theorem applies.\\

\noindent \textbf{Theorem 1.}
Suppose $\sigma \in SU(d)$ and a perturbation $\sigma \rightarrow \rho$ occurs, where $d_{HS}(\sigma, \rho) \leq \epsilon$ and $d_{HS}$ is the Hilbert-Schmidt distance. If the adversarial risk is bounded above by $R$, then $\epsilon^2$ must be bounded above by 
\begin{align}\label{eq:sudepsilon1}
    \epsilon^2<\frac{4}{d} \ln \left(\frac{2}{\mu(\mathcal{M})(1-R)}\right).
\end{align}

\noindent \textit{Proof.} We only give the basic intuition behind the proof here, which is also based on recent classical machine learning results in \cite{mahloujifar2018curse}. Please refer to Appendix~\ref{app:theorem1} for details. It is possible to show that having $SU(d)$ equipped with the Haar probability measure and Hilbert-Schmidt metric belongs to the $(\sqrt{2}, 1/4)-$normal Levy family, thus obeying Eq.~\eqref{eq:normallevy}. We choose $\Sigma=SU(d)$. If the subset of misclassified points is denoted $\tilde{\Sigma}=\mathcal{M}$, then Eq.~\eqref{eq:normallevy} shows that more and more points tend to concentrate around the boundary of $\mathcal{M}$ as $d$ increases. So if we select a point outside $\mathcal{M}$, as $d$ increases, the distance between the chosen point and the nearest point in $\mathcal{M}$ becomes smaller. This makes it more likely that only a small perturbation can result in misclassification. $\square$ 

From theorem 1, we see that with large $d$, even a very small perturbation can result in a misclassification. As a result, it also becomes more and more difficult to certify whether or not $\sigma$ has been adversarially perturbed. Furthermore, to certify whether a small perturbation has occurred requires extra resources in the quantum setting and these resources grow with increasing dimension. Thus a tension develops between the resources required to ensure robustness against misclassification and the quantum advantage expected as the dimension grows.  

Our aim is then to find fundamental limits to this tension that is independent of the details of any particular classification protocol. To describe this trade-off more precisely, in the following sections we examine two general scenarios, which we call untrusted state preparation and untrusted device preparation, depicted in Figure 1.

\begin{figure*}
\includegraphics[width=0.8\textwidth]{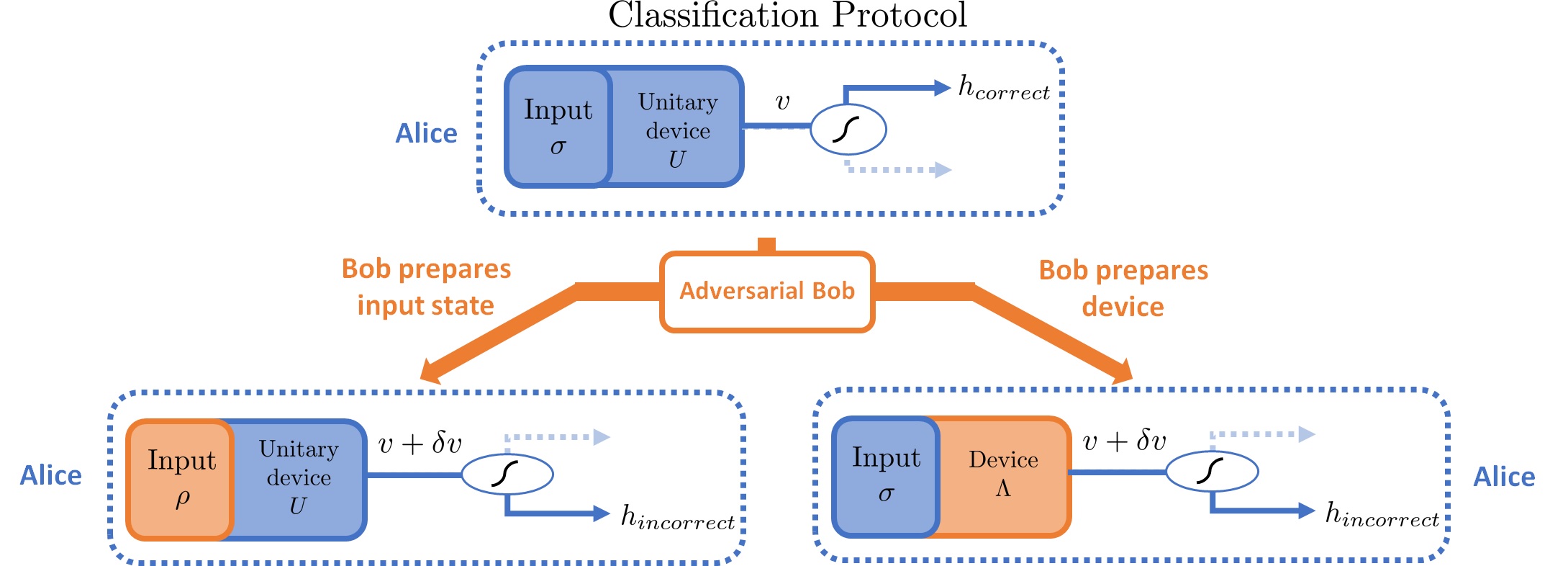}
\caption{\label{fig:Fig1} \textit{Top.} In the absence of an adversarial Bob, Alice performs her classification task in three steps. First she prepares her input quantum state $\sigma$. She then inputs $\sigma$ into her classification device which outputs a real number $v(\sigma)$. Then Alice computes the corresponding class label by applying some thresholding function $v(\sigma) \rightarrow h(\sigma)$ and obtains the correct class label $h_{correct}$. However, in many situations, Alice would want to delegate either her state preparation of $\sigma$ or the preparation of her classification device to Bob. Even if Alice can certify the state or the device after receiving it from an adversarial Bob, she can still misclassify the input state if the dimension of $\sigma$ is high. \textit{Untrusted state preparation (bottom left):} Alice asks Bob for state $\sigma$ and Bob instead returns the state $\rho$. Alice certifies that $\rho$ is close to $\sigma$ in fidelity and gets the output $v(\rho)=v(\sigma)+\delta v(\sigma)$ from her device. Even when $\delta v$ is small, once mapped to $h$, there can be a misclassification $h_{incorrect}$. \textit{Untrusted device preparation (bottom right):} Similarly, suppose Bob prepares the device implementing $\Lambda$ instead of $U$ and gives it to Alice. She is able to certify the device to some precision, and obtains the output $v+\delta v$ where $\delta v$ is small. However, small $\delta v$ can still result in a misclassified label $h_{incorrect}$ if the dimension of $\sigma$ is high enough.}
\end{figure*}
\noindent \textit{Untrusted state preparation---} Alice wants to prepare and classify a $d$-dimensional pure quantum state $\sigma$, or equivalently a $\log_2 d$-qubit state. Let $\sigma=U|b\rangle \langle b|U^{\dagger}$, where $U$ is Haar-randomly selected from $SU(d)$ and $|b\rangle$ is some given pure $d$-dimensional quantum state. Alice has a classical description of her $\sigma$, but does not have the quantum resources to create $\sigma$. She delegates to Bob the preparation of $\mathcal{N}$ copies of $\sigma$. 

However, Bob may be adversarial. He instead prepares a state $\tilde{\rho}$. This has the same dimensionality as $\sigma^{\otimes \mathcal{N}}$ and  $\tilde{\rho}=\sigma^{\otimes \mathcal{N}}$ only when Bob is honest. So Alice assigns $\mathcal{N}_{state}$ registers of $\tilde{\rho}$ to certifying whether Bob could be an adversary. She can choose either state tomography or state certification for her test. Since state tomography demands more resources than state certification, it suffices to examine only the latter.

In general, Bob has the capability of preparing any $\tilde{\rho}$ where (i) $\tilde{\rho}$ is some general entangled $\mathcal{N}$ register state with dimension $2^{\mathcal{N}d}$, (ii) $\tilde{\rho}=\otimes_i^\mathcal{N} \rho_i$ where each $\rho_i$ has dimension $d$ or (iii) $\tilde{\rho}=\rho^{\otimes \mathcal{N}}$. For our purpose here in finding a lower bound to $\mathcal{N}_{state}$, it is sufficient to consider case (iii). Thus Alice's certification task is to ascertain if $F(\rho, \sigma)>1-\chi$ to some high probability $1-\Delta$ where $\chi$ small enough so that adversarial perturbations cannot be hidden. If this condition is not satisfied, then Alice rejects the state. Since (i) and (ii) require larger $\mathcal{N}_{state}$ \footnote{For instance, see \cite{yukiprx} or \cite{Pallister}.} our following results still hold for these cases.

There are two main approaches to 
state certification. The first is the direct estimation of fidelity $F(\rho, \sigma)$. To estimate the fidelity to precision $\eta$ with failure probability $\Delta$ requires the measurement of at least $1/(\Delta\eta^2)$ different Pauli observables. Hence at least this many copies of $\rho$ are needed \cite{flammia}. An alternative approach relies on a collection of binary-outcome measurements used to make the binary decision that the state satisfies either $F(\rho, \sigma)=1$ or $F(\rho, \sigma)<1-\chi$ \footnote{For example, see \cite{Pallister}}. However, since our aim is to find the minimal resource requirements for when $F(\rho, \sigma)>1-\chi$, the fidelity estimation approach better suits our purpose. 

While improvements to state certification methods attempt to decrease the dependence of $\mathcal{N}_{state}$ on $d$, the assumption is always of a constant $\eta$ that is independent of $d$. The key difference in our scenario is that now it is $\eta$ itself that carries a dimensional dependence. If Alice now sets $R$ as the upper-bound to the adversarial risk she is willing to tolerate, the following statement holds true. 

\noindent \textbf{Theorem 2a.} 
Let Alice's quantum classification device implement the function $h(\sigma)$ and let $\mathcal{M}$ be the set of states that $h(\sigma)\neq c(\sigma)$. Alice needs $\mathcal{N}_{state}$ copies of $\rho$ to estimate the fidelity $\mathcal{F}(\sigma, \rho)$ to a precision that is necessary to guarantee that the adversarial risk is at most $R$ with failure probability $\Delta$. Then she requires at least 
\begin{align}
    \mathcal{N}_{state}\geq \frac{d^4}{g^2(\mu(\mathcal{M}), R)\Delta}
\end{align}
copies of $\rho$, where $g(\mu(\mathcal{M}), R)=2\ln(2/(\mu(\mathcal{M})(1-R)))$.  

\noindent \textit{Proof.}  We sketch the basic ideas behind the proof here and refer the reader to Appendix~\ref{app:theorem2a} for details. Alice's selection of state $\sigma$ is equivalent to her selecting $U$ Haar-randomly from $SU(d)$ \footnote{Since global phases do not affect any outcomes of physical measurements, our results here would be equivalent to the case if we used $U(d)$ instead.}. Thus we can use Eq.~\eqref{eq:sudepsilon1} in Theorem 1. Next we need to change the Hilbert-Schmidt distance $\epsilon$ into the corresponding quantum fidelity, used for Alice's state certification scheme. Then Eq.~\eqref{eq:sudepsilon1} can be shown to be equivalent to $F(\sigma,\rho)\geq 1-g(\mu(\mathcal{M}), R)/d^2$ where $g(\mu(\mathcal{M}), R)$ is a function independent of $d$. This means that the error in estimating the fidelity must satisfy $\eta<g(\mu(\mathcal{M}), R)/d^2$. Using the fidelity estimation approach \cite{flammia}, $\mathcal{N}_{state}> 1/(\Delta \eta^2)$. Thus  $\mathcal{N}_{state}>d^4/(g^2(\mu(\mathcal{M}), R) \Delta)$. $\square$

It is crucial to note that now $\mathcal{N}$ has a polynomial dependence on the dimension, which derives from the polynomial dependence of $\eta$ on dimensionality. This has significant implications for the robustness of classification protocols against adversarial perturbations. Theorem 2a states that the minimum certification cost is exponential in the size of the quantum system, even when the certification protocol is considered otherwise efficient. 

For example, in quantum-enhanced machine learning algorithms, there are claims of quadratic and even up to exponential quantum speedups in $d$ compared to the corresponding classical machine learning algorithms \cite{biamonte2017quantum}. However, it is often assumed that the state preparation for converting classical data into quantum input states is delegated to an oracle, otherwise the resources for state preparation itself can over-ride resources potentially saved by employing a quantum algorithm \cite{aaronson2015read}. This could be made possible for instance by delegating quantum-random-access memory (QRAM) capabilities to Bob \cite{biamonte2017quantum}. Yet, we saw that, even in the presence of such an oracle, Alice's certification requirements are of order at least $d^4$. This would then still overwhelm any claimed exponential speedup and many quadratic speedup advantages afforded by these algorithms. From this we see a trade-off between security and speedup as dimensionality increases.

\noindent \textit{Untrusted device preparation---} Suppose Alice now delegates only the preparation of the unitary part $U$ of her classification device to a possibly adversarial Bob. Bob has no other power than the device preparation. If Bob is honest, he returns the prepared $U$ to Alice, she inputs the quantum states $|\Psi\rangle$ whose source she trusts. She then makes the final measurement on her device to obtain the classification of those states. If Bob is adversarial, then he may prepare a device that implements the CPTP map $\Lambda$ instead. Given that Alice knows the classical description of $U$, she is able to perform device certification, by estimating the average channel-fidelity $\text{avg}_{\{|\Psi\rangle\}} \text{Tr}(\Lambda(|\Psi\rangle \langle \Psi|)U |\Psi\rangle \langle \Psi|U^{\dagger})$ \cite{flammia}, where $|\Psi\rangle$ is selected Haar-randomly. Then by using at least $\mathcal{N}_{device}>1/(\Delta' \delta^2)$ calls to the device, Alice can estimate this fidelity to precision $\delta$ with failure probability $\Delta'$. Similarly to the case in untrusted state preparation, if Alice wants to classify the states $|\Psi\rangle$ selected like in Theorem 2a, the following result holds.\\

\noindent \textbf{Theorem 2b.} Let $U$ be the unitary that Alice wants to implement, which is used in a classification device that misclassifies inputs states with probability $\mu(\mathcal{M})$. Suppose Bob prepares, instead of $U$, the circuit implementing a CPTP map $\Lambda$. Then Alice requires at least 
\begin{align}
    \mathcal{N}_{device}> \frac{d^4}{g^2(\mu(\mathcal{M}),R')\Delta'}   
\end{align}
calls to her device to estimate the average channel-fidelity to guarantee that that her adversarial risk is bounded above by $R'$ with failure probability $\Delta'$. 
\\
\noindent \textit{Proof.} The ideas behind the proof are very similar to Theorem 2a. Let $\delta$ now be the error in estimating the average channel-fidelity with failure probability $\Delta'$. The channel estimation protocol \cite{flammia} requires that $\mathcal{N}_{device}>1/(\Delta' \delta^2)$. Again employing Theorem 1 and converting the Hilbert-Schmidt distance to that of average channel-fidelity, we find $\delta< g(\mu(\mathcal{M}), R')/d^2$. Thus the number of extra certification resources required where adversarial risk is bounded above by $R'$ is  $\mathcal{N}_{device}>d^4/(g^2(\mu(\mathcal{M}), R')\Delta')$. For details see Appendix~\ref{app:theorem2b}. $\square$ 

An important example is in the context of quantum learning, where we wish to classify states that we do \textit{not} have the classical descriptions of. These include for instance quantum-template matching \cite{sasaki2002quantum} and quantum anomaly detection \cite{liu2018quantum}, where the control-SWAP gate serves as the key component to the classification device. Since control-SWAP gates, especially for large dimensional systems, are difficult to experimentally realise \cite{patel2016quantum}, these would be the gates delegated to Bob to prepare. However, certifying the control-SWAP gates to a constant precision $\eta$ is insufficient and the minimum $\eta$ to be robust against adversarial attacks must grow with scaling at least $d^2$ \footnote{This differs from the delegated quantum machine learning setting in \cite{sheng2017distributed}, which assumes repeated interactions between Alice and Bob during the computation.}. 

\noindent \textit{Other applications---}It is also possible to apply our results to cases where machine learning is used to identify witnesses of physical properties like non-local correlations or the presence of phase transitions. Our results depend on a crucial requirement that the misclassification probability $\mu(\mathcal{M})>0$. This is almost always true when the classification function is learned from examples rather than derived from first principles. For instance, in ~\cite{ma2018transforming,canabarro2018machine}, machine learning algorithms and training data are used to learn the decision boundary to determine whether a quantum state is entangled or separable and here $\mu(\mathcal{M})>0$. Other examples include learning decision boundaries for classifying phases \cite{carrasquilla2017machine}. These scenarios are all vulnerable to adversarial perturbations described in this paper. 

\noindent \textit{Discussion---} While high-dimensional Hilbert spaces are often considered beneficial for quantum information processing, its vulnerabilities apart from sensitivity to decoherence are almost entirely unexplored. In this paper we examine one important instance. For classification problems involving quantum states or processes, we demonstrate a fundamental bound showing how higher-dimensional quantum systems are more sensitive to adversarial perturbations than lower-dimensional systems. This phenomenon is independent of any decoherence effects or details of any particular classification protocol. Furthermore, any exponential quantum advantages in classification are erased by this behaviour.

Our results so far come from finding bounds to the minimum distance between Alice's original datum $\sigma$ and the nearest element in the originally  misclassified set $\mathcal{M}$. However, these nearest elements to $\sigma$ might have zero measure, thus cannot be reached if Bob perturbs $\sigma$ in random directions. These cases are significant to consider, because it means that these misclassified points can only be reached from $\sigma$ by adversarial perturbations, i.e., Bob deliberately moving $\sigma$ to an element in $\mathcal{M}$. Thus, an important future direction is to examine the differences between the resources required for Alice to remain 
robust against adversarial perturbations to that of random perturbations or perturbations induced by noise, for instance in \cite{cross2015quantum}. This is particularly poignant in the NISQ-era of quantum technologies.

To further understand the effects of high-dimensionality, it is important to go to the extreme case of continuous-variable quantum information, with potentially infinite degrees of freedom. It is then compelling to investigate whether these quantum systems possess some maximal sensitivity to adversaries and if this power can be harnessed to devise novel detection methods. 

We are now only unravelling the beginnings of how high-dimensional quantum systems behave under adversarial perturbations. This has both practical and foundational ramifications. Its relevance will increasingly grow as quantum devices are networked together in a future quantum internet \cite{quantuminternetbook} and adversarial considerations become inevitable. 
%Phase transitions interesting for the study of quantum simulators for quantum computation 

%Analogue quantum simulator\\
%\section{Criticism}
%To include somewhere in text:\\

%$\bullet$ Criticism: Here assume that the test states are drawn from the full set of pure states with equal measure. In real scenarios, there is usually some set distribution that one is choosing from (e.g. due to some priori knowledge), and is \textit{not} completely random. \\

%$\bullet$ Also suggests that better certification protocols are needed that focus on better scaling with precision $1/\eta^2$. \\
\section*{Acknowledgements}
We thank Micha{\l} Oszmaniec (University of Gda\'nsk), Roger Melko (Perimeter Institute for Theoretical Physics and Institute for Quantum Computing, University of Waterloo), Gael Sent\'is (University of Siegen) and Yadong Wu (University of Calgary) for interesting discussions. Our special thanks to Barry Sanders (University of Calgary, University of Science and Technology China) for both fruitful discussions and feedback on the manuscript. 
\bibliography{bibliography}
\appendix 
\section{Robustness of quantum devices against perturbations in the input state} \label{app:uhlmann}
Quantum devices used as part of computational problems are able to provide a real-number output $v(\sigma)$ from a given input quantum state $\sigma$. These devices implement a CPTP map $\Lambda$ and a real number can be extracted by performing quantum measurements on $\Lambda(\sigma)$ to obtain $\text{Tr}(\mathcal{O}\Lambda(\sigma))$, where $\mathcal{O}$ is an observable or POVM. 

%For example, in Deutsch's algorithm or the quantum phase estimation algorithm, direct read-out of the output state provides the result of the problem. For probabilistic algorithms like the SWAP-test, that measures the fidelity between two unknown states, $\Lambda$ is the control-SWAP gate. Or for witness-type scenarios (like measuring an entanglement-witness), the goal is find whether $\text{Tr}(\mathcal{O}\Lambda(\sigma))>0$ or  $\text{Tr}(\mathcal{O}\Lambda(\sigma))<0$.\\

Suppose we allow a perturbation of our initial state $\sigma \rightarrow \rho$ where $\mathcal{F}(\sigma, \rho) \geq 1-\chi$ and $\chi<<1$. Then for a CPTP map $\Lambda$, Uhlmann's theorem \cite{nielsen2000quantum} holds 
\begin{align}
     \mathcal{F}(\Lambda(\rho), \Lambda(\sigma))\geq \mathcal{F}(\sigma, \rho) \geq 1-\chi. 
\end{align}
Thus any small adversarial perturbation of $\sigma$ results in an even smaller perturbation of the final state $\Lambda(\sigma)$. We define $v(\sigma)=\text{Tr}(\mathcal{O}\Lambda(\sigma))$ as the output of the quantum device and assume that both $\mathcal{O}$ and $\Lambda(\rho)-\Lambda(\sigma)$ are positive semi-definite. This means that we can always find matrices $A$, $B$ such that $\mathcal{O}=A^{\dagger}A$ and $\Lambda(\rho)-\Lambda(\sigma)=B^{\dagger}B$. The Frobenius norm of the matrix $AB$ satisfies the norm inequality $||AB||^2_F \equiv \text{Tr}(A^{\dagger}A BB^{\dagger}) \leq ||A||^2_F||B||^2_F=\text{Tr}(A^{\dagger}A)\text{Tr}(B^{\dagger}B)$. This means that $\text{Tr}(\mathcal{O}(\Lambda(\rho)-\Lambda(\sigma)))\leq \text{Tr}(\mathcal{O})\text{Tr}(\Lambda(\rho)-\Lambda(\sigma)))$. Thus we obtain
\begin{align}
    & |v(\sigma)-v(\rho)|\leq |\text{Tr}(\mathcal{O}) \text{Tr}(\Lambda(\sigma)-\Lambda(\rho))| \nonumber \\
    &\leq 2\text{Tr}(\mathcal{O}) \sqrt{1-\mathcal{F}(\Lambda(\sigma), \Lambda(\rho))^2} \nonumber \\
    &\leq 2\text{Tr}(\mathcal{O}) \sqrt{\chi(2-\chi)}< 2\sqrt{2}\text{Tr}(\mathcal{O}) \sqrt{\chi},
\end{align}
where we used the Fuchs–van de Graaf inequality for the trace distance $|\text{Tr}(\Lambda(\sigma)-\Lambda(\rho))|\leq 2\sqrt{1-F(\Lambda(\sigma),\Lambda(\rho))^2}$ in the second line. 

This means that, if $\text{Tr}(\mathcal{O})$ is small, then a small $\chi$ corresponds to a small difference to the desired outcome $v(\sigma)$. 

However, for classification problems where the final output is constrained to a finite number of possible values, we will see that even small $|v(\sigma)-v(\rho)|$ can result in misclassification. 
%Note for later: what if Tr(O) scales with d? Here we are assuming this is NOT the case. 
\section{Proof of Theorem 1}
\label{app:theorem1}
Let us begin with some definitions. We start with a normalised metric probability space which is the set $\Sigma$ equipped with a probability measure $\mu$ and a metric $D$, where $\mu(\Sigma)=1$. For any $\tilde{\Sigma}\subset \Sigma$, $\mu(\tilde{\Sigma})$ can be interpreted as the probability that some point in $\tilde{\Sigma}$ is selected. Every $\sigma \in \Sigma$ is assigned a class label from a countable set, which is determined by the function $h(\sigma)$.  

For $\tilde{\Sigma}\subset \Sigma$ with distance measure $D$ and probability measure $\mu$, we can define the concentration function $\alpha(\epsilon)$ as
\begin{align} \label{eq:concentration}
    \alpha(\epsilon) \equiv 1-\inf \{\mu(\tilde{\Sigma}_{\epsilon})|\mu(\tilde{\Sigma})\geq 1/2\},
\end{align}
where $\tilde{\Sigma}_{\epsilon}$ is the $\epsilon-$expansion of $\tilde{\Sigma}$ defined in Eq.~\eqref{eq:epsilonexp} in definition 3. Here $\epsilon$ is measured with respect to the distance measure $D$. The concentration function quantifies the extent to which points in $\Sigma$ concentrate about boundary lines in $\Sigma$ and this plays a crucial role in isoperimetric theorems.

For our purpose where $\Sigma$ is the special unitary group $SU(d)$, and the corresponding metric probability space belongs to a family of spaces with $SU(j)$ where $j$ is a positive integer, including $j=d$. Such a metric probability space is said to be a $(l_1,l_2)-$normal Levy family if their corresponding concentration function $\alpha (\epsilon)$ is bounded above by
\begin{align}
    \alpha(\epsilon)\leq l_1 e^{-l_2\epsilon^2d}.
\end{align}
Suppose $\sigma$ is correctly classified, i.e., $h(\sigma)=c(\sigma)$. Our aim is then to find the smallest size $\epsilon$ of the perturbation $\sigma \rightarrow \rho$ so that  $D(\sigma,\rho)\geq \epsilon$ and $\rho$ is misclassified with adversarial risk at most $R$ (see Definition 3). This would be an adversarial perturbation that is small enough so it is expected $c(\sigma)=c(\rho)$, but the class label computed for $\rho$ differs from $\sigma$, i.e.,  $h(\rho) \neq c(\rho)=c(\sigma)$.

This is equivalent to finding the shortest distance
between $\sigma$ and some point in the set of misclassified points $\mathcal{M}=\{\sigma \in \Sigma|h(\sigma)\neq c(\sigma)\}$. If this distance $\epsilon$ satisfies 
\begin{align} \label{eq:mahmoody}
    \epsilon^2 > \frac{1}{l_2d}
(\ln(l_1/\mu(\mathcal{M}))+\ln(l_1/\gamma)),
\end{align} 
then from Theorem 3.7 in \cite{mahloujifar2018curse}, it can be  shown that the adversarial risk is $R_{\epsilon}\geq 1-\gamma$. For completeness, we present our condensed version of the proof below. 

Suppose we define $\epsilon_1$ where $\mu(\mathcal{M})> l_1 \exp(-l_2\epsilon^2_1d)$. Let the $\epsilon_1-$expansion of $\mathcal{M}$ be denoted $\tilde{\mathcal{M}}$. Then $\mu(\tilde{\mathcal{M}})>1/2$. We can see this by looking at cases (i) $\mu(\mathcal{M})>1/2$ and (ii) $\mu(\mathcal{M})<1/2$ separately. When (i) is true,  $\mu(\tilde{\mathcal{M}})>1/2$ follows immediately from $\mu(\tilde{\mathcal{M}})>\mu(\mathcal{M})$. For (ii), suppose $\mu(\tilde{\mathcal{M}})<1/2$. Then $\mu(\Sigma \setminus \tilde{\mathcal{M}})\geq 1/2$. The set $\tilde{\Sigma}=\Sigma \setminus \tilde{\mathcal{M}}$ then gives rise to a corresponding concentration function $\alpha(\epsilon_1)\geq 1-\mu(\tilde{\Sigma}_{\epsilon_1})$ where $\alpha(\epsilon_1)<\mu(\mathcal{M})$. Thus $\mu(\mathcal{M})+\mu(\tilde{\Sigma}_{\epsilon_1})\geq 1$, which means the two sets $\mathcal{M}$ and $\tilde{\Sigma}_{\epsilon_1}$ share a point. However, this also means that there is a point that is simultaneously in $\tilde{\mathcal{M}}$ and $\tilde{\Sigma}$, which is a contradiction. Therefore, $\mu(\tilde{\mathcal{M}})>1/2$. 

Now we can look at the  $\epsilon_2-$expansion of $\tilde{\mathcal{M}}$, which is equivalent to the $(\epsilon_1+\epsilon_2)-$expansion of $\mathcal{M}$. Then the corresponding concentration function is $\alpha(\epsilon_2)\equiv 1-\inf \{\mu(\tilde{M}_{\epsilon_2})|\mu(\tilde{\mathcal{M}}\}$. Thus the adversarial risk $R_{\epsilon}$ associated with a perturbation by $\epsilon=\epsilon_1+\epsilon_2$ is $\mu(\mathcal{M}_{\epsilon_1+\epsilon_2})=\mu(\tilde{\mathcal{M}}_{\epsilon_2})\geq 1-\alpha(\epsilon_2)$. If we now define $\gamma=l_1\exp(-l_2\epsilon_2^2d)\geq \alpha(\epsilon_2)$, then $R_{\epsilon}\geq 1-\gamma$. 

Now we can set an upper-bound $R$ to the adversarial risk $ R_{\epsilon} \leq R$. This is the maximum tolerated probability of misclassification. By making the replacement $\gamma=1-R$ in Eq.~\eqref{eq:mahmoody}, we find that if  $1-R>(1/\mu(\mathcal{M})) l_1^2\exp(-\epsilon^2 l_2 d)$, then $R_{\epsilon}\geq R$. However, this is a contradiction of our condition that $R_{\epsilon} \leq R$. Thus, in order to satisfy $R_{\epsilon} \leq R$, a necessary condition to satisfy is  $1-R<(1/\mu(\mathcal{M})) l_1^2\exp(-\epsilon^2 l_2 d)$.

This then results in our following theorem:  A necessary condition for the adversarial risk to be bounded above by $R_{\epsilon} \leq R$ is 
\begin{align} \label{eq:mastereq}
    \epsilon^2<\frac{1}{l_2 d} \ln \left(\frac{l_1^2}{\mu(\mathcal{M})(1-R)}\right).
\end{align}
The interpretations here are quite clear: (i) If the dimensionality $d$ is high, then a much smaller perturbation can result in misclassification; (ii) If $R$ is low, then a smaller perturbation is sufficient to result in misclassification; (iii) If the risk $\mu(\mathcal{M})$ is initially high, then only a small perturbation can cause misclassification; (iv) If $l_1^2$ is low, then $\epsilon$ is bounded to be small. This is saying that if the space is more \textit{concentrated} (i.e., more area is covered by a small $\epsilon$-expansion), then the adversarial behaviour is more pronounced.

We now focus our attention on when $\Sigma$ is the group $SU(d)$ and is equipped with the normalised Hilbert-Schmidt metric and the Haar probability measure. We now show this metric probability space is a $(\sqrt{2}, 1/4)$-normal Levy family \cite{gromov1983topological, giordano2007some}. We first employ the isoperimetric inequality \cite{milman2009asymptotic, gromov1983topological, gromov1980paul}, which states that for a subset $\tilde{\Sigma} \subset \Sigma$, where $d=\dim \Sigma$, and $\mu(\tilde{\Sigma})\geq 1/2$, the measure of the $\epsilon$-expansion of $\tilde{\Sigma}$ satisfies
\begin{align} \label{eq:isop}
    \mu(\tilde{\Sigma}_{\epsilon})\geq 1-\sqrt{2}e^{-\epsilon^2 d R(\Sigma)/(2(d-1))}.
\end{align}
Here $R(\Sigma)=\inf_{\tau} \text{Ric}(\tau, \tau)$ and  $\text{Ric}(\tau, \tau')$ is the Ricci curvature of $\Sigma$, where $\tau$ runs over all unit tangent vectors. The condition for the isoperimetric inequality to hold requires $R(\Sigma)=(d-1)/r^2$, which is the Ricci scalar of a $d$-dimensional sphere with radius $r$. Then $\mu$ is the normalised Haar measure on this $d$-dimensional sphere with radius $r$. 

Using the definition of the concentration function in Eq.~\eqref{eq:concentration},  Eq.~\eqref{eq:isop} then implies
\begin{align} \label{eq:sudalpha}
    \alpha(\epsilon) \leq \sqrt{2}e^{-\epsilon^2 d R(\Sigma)/(2(d-1))}.
\end{align}
From \cite{meckes2014concentration}, it can be shown for $SU(d)$ that $\text{Ric}(\tau, \tau)=\frac{d}{2} G(\tau, \tau)$, where $G(\tau, \tau)$ is the Hilbert-Schmidt metric and $\tau$ is any vector in the tangent space of $SU(d)$. Therefore, $R(\Sigma)=(d/2)\inf_{\tau} G(\tau, \tau)$. Then from \cite{oszmaniec2016random} we find that since $G(\tau, \tau)=1$, $R(X)=d/2$. 

This means for $SU(d)$ we can rewrite Eq.~\eqref{eq:sudalpha} as 
\begin{align} \label{eq:sudalpha}
    \alpha(\epsilon) \leq \sqrt{2}e^{-\epsilon^2 d^2/(4(d-1))} < \sqrt{2}e^{-\epsilon^2 d/4},
\end{align}
thus showing $SU(d)$ with the Haar measure and the Hilbert-Schmidt metric is a $(\sqrt{2}, 1/4)-$normal Levy family. 
%potential change: d-> d^2-1
Then inserting $l_1=\sqrt{2}$ and $l_2=1/4$ into Eq.~\eqref{eq:mastereq}, we find 
\begin{align} \label{eq:sudepsilon}
    \epsilon^2<\frac{4}{d}\ln \left(\frac{2}{\mu(\mathcal{M})(1-R)}\right),
\end{align}
where $\epsilon$ is measured in terms of the Hilbert-Schmidt distance. 
\section{Proof of Theorem 2a}
\label{app:theorem2a}
Suppose we begin with a $d$-dimensional pure state $|b\rangle$. Then by randomly selecting a unitary $u_i \in SU(d)$, this is equivalent to selecting a random pure state defined by $|\psi_i\rangle \equiv u_i|b\rangle$. %\footnote{In fact, it can be shown that the Haar randomness for the set of unitaries has a one-to-one relationship to Haar randomness for pure quantum states [REF].}. 
Then we can map the problem of classifying pure states to the problem of classifying unitary operations in $SU(d)$, which satisfies the concentration inequalities we have found in Theorem 1. Note that here we can work with $u_i\in SU(d)$ instead of $U(d)$ since global phases do not make measurable differences to the states.\\

In Theorem 1, our distance $\epsilon$ is in terms of the Hilbert-Schmidt distance between two unitaries $u_1$, $u_2$, which is defined as
\begin{align}
   & H(u_1, u_2)^2=\text{Tr}((u_1-u_2)^{\dagger}(u_1-u_2))\nonumber \\
   &=2d-\text{Tr}(u^{\dagger}_1u_2+u^{\dagger}_2u_2).
\end{align}
Let $\{|b_i\rangle\}$ be a basis set in our $d$-dimensional Hilbert space. Then the Hilbert-Schmidt distance satisfies the following inequality
\begin{align} \label{eq:hepsilon}
  & H(u_1, u_2)^2=2d-\sum_{i=1}^d \langle b_i|u^{\dagger}_1u_2+u^{\dagger}_2u_1|b_i\rangle \\ \nonumber
  &\geq 2d-d \max_i \langle b_i|u^{\dagger}_1u_2+u^{\dagger}_2u_1|b_i\rangle \\ \nonumber
  &\geq 2d-d \langle b|u^{\dagger}_1u_2+u^{\dagger}_2u_1|b\rangle \nonumber \\
  & \geq 2d-d (\langle \psi_1|\psi_2\rangle +\langle \psi_2|\psi_1\rangle)=2d(1-\Re\langle \psi_1|\psi_2\rangle) \nonumber \\
  &\geq 2d(1-F(|\psi_1\rangle, |\psi_2\rangle)),
\end{align}
where $F(|\psi_1\rangle, |\psi_2\rangle)\equiv |\langle \psi_1|\psi_2\rangle|$ is the quantum fidelity between two pure states. Thus we have related the Hilbert-Schmidt distance between two unitaries to a distance-like measure between two quantum states. This inequality is also tight, since $H(u_1,u_2) \rightarrow 0$ as $F(u_1,u_2)\rightarrow 1$.

We can now replace $\epsilon$ in Eq.~\eqref{eq:sudepsilon1} with $H(u_1, u_2)$. Suppose $|\psi_1\rangle$ is the state Alice wants Bob to prepare, but Bob instead prepares $|\psi_2\rangle$. Then inserting Eq.~\eqref{eq:hepsilon} into Eq.~\eqref{eq:sudepsilon1} we find that the minimum fidelity between $|\psi_1\rangle$ and the perturbed state $|\psi_2\rangle$ state must be bounded below by
\begin{align} \label{eq:fidelitylower}
    F(|\psi_1\rangle, |\psi_2\rangle) \geq 1-\frac{2}{d^2}\ln\left(\frac{2}{\mu(\mathcal{M})(1-R)}\right),
\end{align}
where $R$ is the upper-bound to the adversarial risk allowed by Alice. We see that $F(|\psi_1\rangle, |\psi_2\rangle) \rightarrow 1$ as $d\rightarrow \infty$. 

Note that the lower bound in Eq.~\eqref{eq:fidelitylower} still holds even when Bob is allowed more general perturbations beyond the purely unitary adversarial operations $|\psi_1\rangle \rightarrow |\psi_2\rangle$, since more general perturbations must also include unitary operations.

Using the direct fidelity estimation protocol in \cite{flammia}, we know that the number of copies of $\rho$ requires is at least
\begin{align} \label{eq:Nstate}
    \mathcal{N}_{state}> \frac{1}{\eta^2 \Delta}.
\end{align}
In fact, since $1/(\Delta \eta^2)$ is only the number of different Pauli observables that need to be measured, the real $\mathcal{N}_{state}$ is in general much greater than this quantity. 

While most work on fidelity estimation focuses on reducing the number of required Pauli observables to measure by careful choices of observables, our focus is on the dependence of $\mathcal{N}$ on $\eta$. It is only from the $\eta$ dependence on $d$ that we have our extra resource cost to protect Alice from adversarial perturbations of $\sigma$. To resolve the value of fidelity to the required precision, from Eq.~\eqref{eq:fidelitylower} we must therefore have 
\begin{align}
    \eta < \frac{2}{d^2}\ln\left(\frac{2}{\mu(\mathcal{M})(1-R)}\right).
\end{align}
Inserting the inequality above into Eq.~\eqref{eq:Nstate}, we find
\begin{align} \label{eq:Nstatefinal}
    \mathcal{N}_{state} > \frac{d^4}{ 4\ln^2 \left(\frac{2}{\mu(\mathcal{M})(1-R)}\right)\Delta}.
\end{align}
 We note that although we could alternatively perform a certification procedure via state tomography, the dependence $\mathcal{N}_{state}\sim 1/\eta$ holds. Thus the resource overhead polynomial in $d$ required in the presence of adversarial perturbations still applies. 
 %Checking the correctness of the theorem: We can also see that as $\mu(\mathcal{M}$ increases or $\Delta$ and $R$ decreases, we require a larger $\mathcal{N}_{state}$.
\section{Proof of Theorem 2b}
\label{app:theorem2b}
In this scenario, we trust the source of the input quantum states, but we do not necessarily trust the quantum device which actually performs the computation. This is very similar to the untrusted state preparation scenario, but now we move the adversarial perturbation from the state to that of the device.

Let Alice trust the input states $\sigma=|\Psi\rangle \langle \Psi|$ to the classification device. Let $U$ denote the unitary gate Alice wants Bob to prepare as part of the classification device. Bob instead prepares some circuit that implements $\Lambda$. For instance, let $UU_{adv}$ denote one possibility of Bob's adversarial perturbation where $U_{adv}$ is another unitary. To guarantee
\begin{align}
    &\text{avg}_{|\Psi\rangle}|\langle \Psi|U^{\dagger} U U_{adv}|\Psi\rangle| \nonumber \\
    &=\text{avg}_{|\Psi\rangle}|\langle \Psi|U_{adv}|\Psi\rangle|\geq 1-\chi'
\end{align}
for some $\chi'$ with failure probability $\Delta'$, one requires an estimation of $\text{avg}_{|\Psi\rangle}|\langle \Psi|U^{\dagger} U U_{adv}|\Psi\rangle|$ to precision $\delta<\chi'$ with the same failure probability. The channel certification protocol in \cite{flammia} can be employed  where $\mathcal{N}_{device}>1/(\Delta' \delta^2)$ calls to the device are needed. 

Similarly to the results in Theorem 2a, a condition that can guarantee robustness against adversarial perturbations with adversarial risk at most $R'$ requires classification device satisfies
\begin{align}
    \min_{|\Psi\rangle}|\langle \Psi|U_{adv}|\Psi\rangle| \geq 1-g(\mu(\mathcal{M}),R')/d^2,
\end{align}
where $g(\mu(\mathcal{M}),R')=2\ln(2/(\mu(\mathcal{M})(1-R')))$. This means that the precision $\delta$ to which we must measure the average channel-fidelity must likewise bounded above by 
\begin{align}
    \delta < g(\mu(\mathcal{M},R'))/d^2.
\end{align}
Thus
\begin{align}
    \mathcal{N}_{device}> \frac{d^4}{g^2(\mu(\mathcal{M},R))\Delta'}.
\end{align}

%%%%%%%%%%%%%%%%%%%%%%%%%%%%%%%%%%%%%%%%%%%%%%%%%%%%%%%%%%%%%%%%%%%%%%%%%%%%%%%

\end{document}